\documentclass{cscw2012}
\pdfoutput=1
\fontfamily{arial}\selectfont
\usepackage{cite}
\usepackage[absolute]{textpos}
\usepackage{graphicx}
\usepackage{amsmath}
\usepackage{url}

\usepackage{times}
\usepackage{balance}

\title{Visualizing Communication on Social Media}
\subtitle{Making Big Data Accessible}
\numberofauthors{1}
\author{
\alignauthor{Karissa McKelvey, Alex Rudnick, Michael D. Conover, Filippo Menczer\\}
\affaddr{Center for Complex Networks and Systems Research \\ 
Indiana University School of Informatics and Computing\\} 
\email{\{krmckelv, alexr, midconov, fil\}@indiana.edu}
}

\makeatletter
\let\@copyrightspace\relax
\makeatother
\begin{document}
\maketitle

\begin{abstract}
The broad adoption of the web as a communication medium has made it possible to
study social behavior at a new scale. With social media networks such as
Twitter, we can collect large data sets of online discourse.
Social science researchers and journalists, however, may not have tools
available to make sense of large amounts of data or of the structure of
large social networks.  In this paper, we describe our recent
extensions to Truthy, a system for collecting and analyzing political discourse
on Twitter.  We introduce several new analytical perspectives on online discourse
with the goal of facilitating collaboration between individuals in the computational and social sciences.  
The design decisions described in this article are motivated by real-world use cases developed in collaboration 
with colleagues at the Indiana University School of Journalism.
\end{abstract}




\section{Introduction}

Online social networking platforms provide a rich and detailed picture of complex sociological phenomena at multiple scales. Recent studies have demonstrated that digital trace data can be combined with sophisticated statistical tools to produce insights into the behavior and interactivity patterns of hundreds of thousands of individual actors~\cite{lazer}. Despite these advances, computational techniques such as machine learning and large-scale network analysis frequently remain beyond the reach of scholars trained in the fields of communication and social science. Consequently, much of the research on sociotechnical systems suffers a lack of theory-driven insight, and is often limited to descriptive accounts of the phenomena under study. Clearly there exists a need for technologies which bridge the gap between quantitative and qualitative means of understanding of social systems.

We argue that interactive visualization tools are a natural solution to this
problem, as they surface large volumes of information about a system, allowing
users to make use of human vision in the development of theory-driven insight.
Information visualization pioneer Ben Schneiderman identifies several key
features of an information visualization tool. Such a tool should allow
users to gain an {\em overview} of the data under study, provide {\em zoom} \&
{\em filtering} capabilities, item-level {\em details-on-demand}, allow users
to see {\em relationships} among items in a collection, and {\em extract}
target data about specific subsets within the collection~\cite{shneiderman1996eyes}.

To this end, we introduce an interactive dashboard for the study of communication processes on the Twitter microblogging platform. Based on the computational architecture of {\em Truthy}\footnote{\url{http://truthy.indiana.edu}}, a system designed to facilitate the tracking and detection of coordinated political deception campaigns, we have created a platform to address the information needs of researchers in the social sciences by providing real-time, interactive visualizations of information diffusion processes on Twitter. Key features of this tool include the ability to produce high-level statistical and visual {\em overviews} of large-scale communication networks; {\em filter} for discussions about specific topics; visualize {\em relationships} among users in core communication networks; leverage statistical inferences about users' influence, demographics and affect; and examine and {\em extract details} about individuals and collections of users based on a variety of filtering criteria.  

\section{The Truthy Platform} 

The \emph{Truthy} system was originally designed to analyze and detect the
emergence of coordinated misinformation campaigns on
Twitter~\cite{politicalabuse,polarization}. Now tasked with the study of
information diffusion in general, Truthy monitors a real-time, high-throughput
feed of 140-character messages known as tweets. Truthy clusters tweets into groups of related messaged called `memes.'
Memes typically correspond to discussion topics, communication channels, or
information resources shared among Twitter users. Here we outline the criteria
for the grouping of content into memes:

\begin{description}

\item[Hashtags] \hfill \\
Hashtags are tokens used to identify the topic or intended audience of a tweet.  For example, {\tt \#taxes} or {\tt \#occupy}.

\item[Mentions] \hfill \\
A Twitter user can include another user's screen name in a post, prefixed with 
the @ symbol. These ``mentions" are used to denote that a particular Twitter
user is being discussed or to address a post to that user.

\item[Hyperlinks] \hfill \\
We extract URLs from tweets by matching strings of valid URL characters that
begin with \texttt{http://}. 

\item[Phrases] \hfill \\
Finally, we consider the entire text of the tweet itself to be a meme once
all Twitter metadata, punctuation, and URLs have been removed. Substrings of
tweets may also be matched. 

\end{description}

As such, we define each meme as the set of all tweets containing a common hashtag, mentioned username, hyperlink, or substring. 
Using these signifiers as the atomic units of information transfer, we are able to create large-scale, high-resolution models of information propagation dynamics. Though we refer the reader to previous works for a detailed description of the Truthy system, here we give a brief overview of some of the key analytical elements afforded by the original framework. 

\begin{figure}
\includegraphics[width=0.5\textwidth]{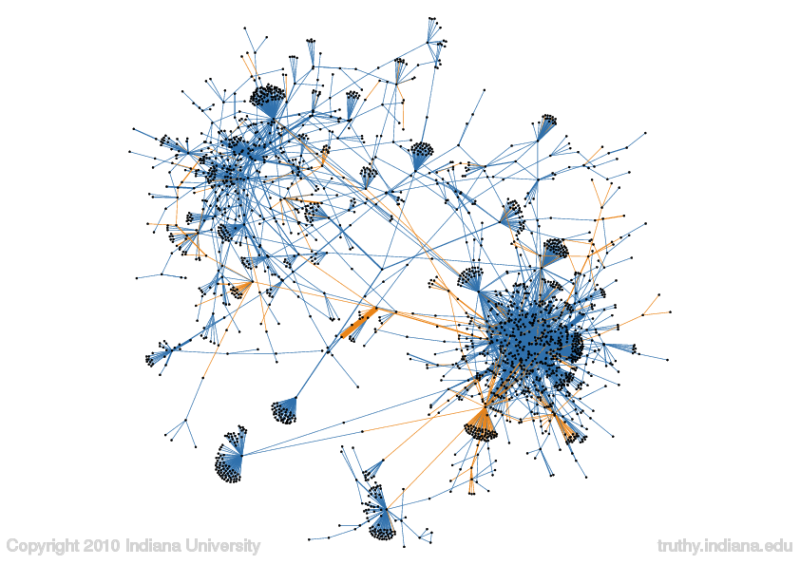}
\caption{Diffusion network associated with the \#usa hashtag. Nodes represent individual users and edges represent two types of tweets, either mentions (orange) or retweets (blue).}
\label{fig:diffusion}
\end{figure}

For ease of navigation, collections of related memes are algorithmically
grouped into top-level categories (`themes')  representing the most
coarse-grained level of analysis available on the Truthy platform. Within a
given theme, users can search for memes containing specific keywords or sort
content based on a variety of statistical features. Navigating a theme, users
are presented with a concise visual representation of each meme, characterized
in terms of a multiplex information diffusion
network~(Figure~\ref{fig:diffusion}) and sparklines representing activity levels over time. 

At the level of an individual meme the user is presented with a high-resolution image of the meme's information diffusion network and a variety of statistics about the activity and connectivity of users who have produced content associated with that meme. These features include the number of users and tweets, diffusion network statistics such as mean degree and large connected component size, and user-specific statistics such as the most retweeted user and the number of unique injection points. Additionally, users can interact with a zoomable historical timeseries of activity volume and produce animations of relevant meme-meme co-occurrence patterns.  

Together these interface elements provide multiple, complementary perspectives on the activity associated with clusters of related content.  Despite the benefits associated with the approach outlined above, this design is largely geared towards providing high-level characterizations of each meme's structure and dynamics. Often, these broad accounts of the system's behavior are inappropriate for users trained in the qualitative analysis of social dynamics. To address this shortcoming, we propose a series of design affordances targeting users who seek to examine the data in an interactive fashion. In doing so, we hope to provide a useful case study for researchers interested in developing technologies that bridge the methodological divide between qualitative and quantitative epistemologies.

\section{Design Goals}

The target users of the system we describe include social scientists, political
scientists, journalists and researchers engaged in the study of communication.
In order to clarify the requirements of this target audience, we worked with
colleagues at the Indiana University School of Journalism to identify use cases
that might be typical of an individual using this system for research purposes.
Below we describe a series of prototypical research questions that motivate the
design of these visualization interfaces.

\begin{enumerate}
\item Are there well-defined communication behaviors that characterize the activities of influential actors?
\item What is the role of bridging users in facilitating information transfer between ideologically opposed communities?
\item Which Twitter accounts act as opinion leaders, and how do they engage in frame-making and agenda-setting?
\end{enumerate}

These examples evoke a number of common themes that help to clarify the
requirements of our prototypical user. At the meme-wide level, users may tend
to be interested in the structural positions occupied by individual actors in a
given communication network.  At the individual level, our target audience
requires fine-grained information, including access to measures of influence,
affect and ideology, information on users' communication choices, and access to
detailed historical data on content production.

To address these requirements, we created interfaces containing several key
analytical components. These elements include an interactive layout of the
communication network shared among a meme's most retweeted users and detailed
user-level metrics on activity volume, sentiment, inferred ideology, language,
communication channel choices, and a real-time feed of each individuals recent
activity.  Additionally, we provide a filterable, searchable, and sortable
table-based interface that allows researchers to make rapid comparisons between
the statistical attributes associated with arbitrary sets of actors in a given
meme. Finally, we provide a novel mechanism to facilitate the acquisition of user- and
meme-level tweet content in a way that falls within the activities permitted of the
Twitter Terms of Service.

\section{Interface Elements}

\subsection{Overview of Calculated Metrics}
\label{sec:metrics}
Here we provide an overview of the data users can expect to encounter while using this system to explore a specific meme. For a finite subset of high-profile users we compute a number of descriptive statistics including their total tweets, retweets and mentions, probable language, affective sentiment with respect to the meme, date of most recent activity, account creation date, and for users engaged in discussions about U.S. Politics, their inferred partisan affiliation.

Sentiment is calculated using OpinionFinder, a system that performs
coarse-grained subjectivity analysis by searching for substrings  in a text to
identify phrases that express positive or negative sentiments
~\cite{opinionfinder}. We also attempt to identify the dominant language of
each Twitter user with the Compact Language Detector
library\footnote{\url{http://code.google.com/p/chromium-compact-language-detector/}},
which was developed by Google for use in the Chrome web browser. This library
makes use of character-level n-grams to estimate the language of a given text.
This may surface useful information; as Twitter is used worldwide, we have
anecdotally observed communities clustering along language boundaries.

To infer the partisan leanings of individual users we apply machine-learning
techniques that leverage network and text features to make 
predictions of political ideology. Originally developed by Conover {\em et
al.}, this technique relies on the highly-clustered structure of a network of
political retweets and the hashtag  usage of individuals in these clusters.
Using a data set of 1,000 manually-annotated users, the authors found that
membership in a specific network cluster accurately predicted users’ political
affiliation with 87.3\% accuracy.  Additionally, the authors found that an
individual's hashtag choices could be used to correctly predict political
affiliation with 83.5\% accuracy~\cite{alignment}.

Here, we combine these two approaches to build a classification model that
allows for the prediction of the political affiliation of hundreds of thousands
of individuals. To this end we relied on a corpus containing all tweets
produced by the 18,470 individuals in either of the two retweet network
communities identified in~\cite{alignment}.  After extracting hashtags from these tweets, we trained a support
vector machine to classify users as belonging to one of the two politically
homogeneous network clusters. Using training and testing sets of 925 and 935 distinct
users, respectively, randomly selected from the network of political retweets we report that
this SVM correctly predicts the cluster association of each user with 85.0\%
accuracy.  Consequently, assuming that network cluster membership is predictive
of political affiliation in 87.3\% of cases (as demonstrated
in~\cite{alignment}), we report an estimated overall accuracy of 85.0*87.3=74.5\%.
We report the ``partisanship" of each user in terms of the 
confidence of the SVM-based prediction, as measured by the distance of the user vector from the
classification hyperplane. For users below some tunable confidence threshold $\epsilon$,
we refrain from reporting a prediction. While a complete analysis of this
methodology is beyond the scope of this article, we propose that this inference
may simply act as a supplementary guide for users of the system, and does not
replace prudent case-by-case judgements.

\subsection{Interactive Meme Diffusion Network}

\begin{figure*}
\centerline{\includegraphics[width=\textwidth]{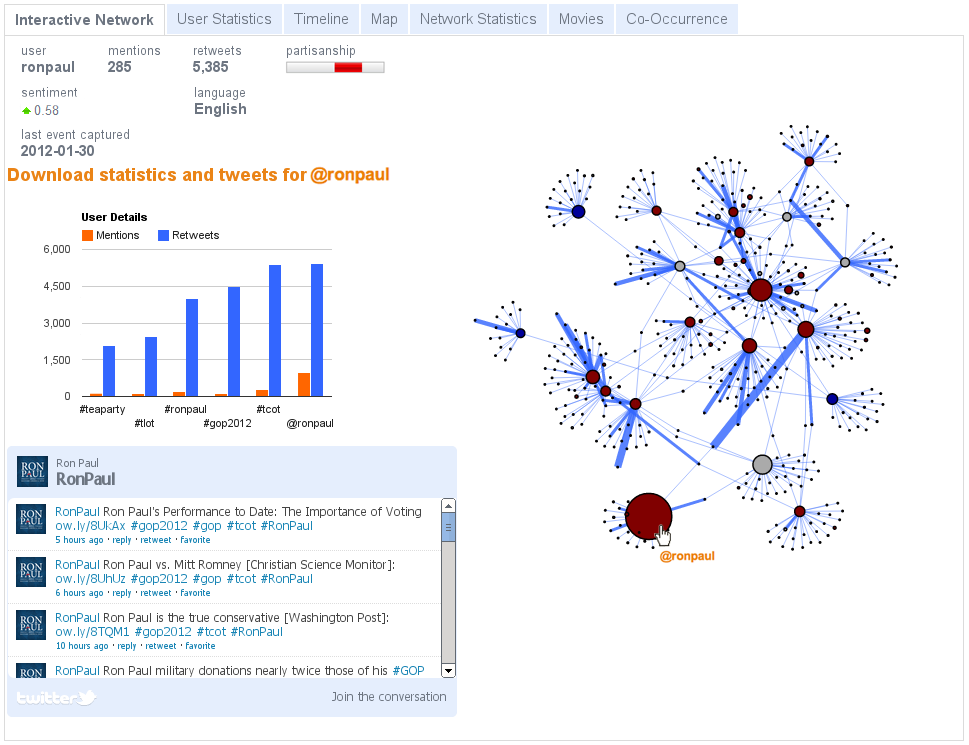}}
\caption{Interactive meme diffusion network visualization and user data exploration interface for content associated with user @ronpaul.}
\label{outdegree}
\end{figure*}

Visible in the right-most portion of Figure~\ref{outdegree}, the meme diffusion network
visualization addresses a key shortcoming of previous work by allowing users to
identify the accounts associated with specific nodes in the communication
network. As many of the motivating research questions dealt with the role of influential users in these communication networks, we filter the original layout (Figure
\ref{fig:diffusion}) to include only the top twenty most retweeted users and their neighbors. This
promotes visual clarity and allows for a detailed examination of the
relationships between the individuals our algorithm estimates to be influential in terms of broadcast reach.

Edges between nodes in this network represent at least one retweet event, and
the width of the line increases with the number of retweets between each pair
of users. Nodes in this network have an area that scales logarithmically with
out-degree, and in the context of themes about U.S. Political Discourse, take a
fill color according to their inferred partisan affiliation.
As per the requirements outlined above, a user can hover over any node in the
network to see the associated account name; clicking a node brings up an
interface detailing a number of features about that user.

The interactive meme diffusion network is created using \emph{d3.js}, a JavaScript library that binds arbitrary data to a Document Object Model (DOM), and then applies transformations using Scalable Vector Graphics~\footnote{\url{http://mbostock.github.com/d3}}.

\subsection{User Data Exploration Interface}

\begin{figure*}
\includegraphics[width=\textwidth]{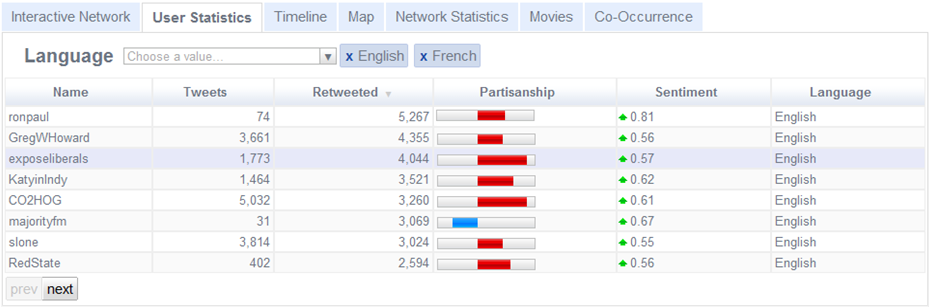}
\caption{Filterable, sortable, searchable data table for users associated with the {\tt \#tcot} meme, a conservative communication channel.}
\label{fig:table}
\end{figure*}

In addition to the data exploration framework described above, we provide a
filterable, searchable, and sortable table based on the Google Chart Tools API
~\footnote{\url{http://code.google.com/apis/chart/interactive/docs/gallery/table.html}}. In this interface (Figure \ref{fig:table}), a data table is presented containing fields corresponding to several of the properties described in the
calculated metrics section. This interface allows
researchers to investigate the behavioral and demographic characteristics of
substantially larger collections of individuals compared to the Interactive
Meme Diffusion Network. For a given set of search criteria, the entire
statistical and demographic information contained in this data table can be
exported as comma-separated values for further study with other analysis
platforms. 

\subsection{Exporting Data and Statistics to Truthy Users}

The Twitter Terms of Service prohibit services that provide direct access to
historical tweet content, instead requiring that requests for data be made
through the official Twitter API.  One of the key demands of our users is the
ability to access and export historical tweet data for individuals and
collections of users. Consequently, we developed a novel approach to this problem
that allows users to download Twitter content by means of client-side API
calls. To this end, we generate locally-executable JavaScript that
automatically downloads tweets corresponding to specific memes or sets of users from the Twitter API.  These data are available in a variety of popular file formats, such as
\texttt{.csv} or Gephi network files. 

\section{Related Work}

In \texttt{The Guardian}'s recent investigation of the UK riot rumors
\footnote{\url{http://www.guardian.co.uk/uk/interactive/2011/dec/07/london-riots-twitter}},
each meme was identified by hand, expanded by a parametrized Levenshtein
distance algorithm, and then independently coded by three sociology PhD
students. These visualizations are helpful in the study of the spread of
misinformation; however, this approach requires significant human labor and
thus is more difficult to scale.

Twitinfo \footnote{\url{http://www.twitinfo.csail.mit.edu}} is a website presenting
research on network analysis and visualizations of Twitter data. Its content is
collected in automatically identified ``bursts" of tweets ~\cite{twitinfo}.
Twitinfo also calculates the top tweeted URLs in each burst, and plots each
tweet on a map, colored according to sentiment.  Twitinfo focuses on specific
memes, identified by the researchers, and is thus somewhat limited for users
who might wish to investigate arbitrary topics.

Ripples is a feature of Google's social network, useful for visualizing the
spread of posts among users; Google+ has a reposting mechanism similar to
Twitter's retweeting. When a user reposts content, Ripples tracks the
intermediate users along the diffusion path, information which is not available
via the Twitter API. Users are represented as colorful bubbles, which
recursively contain the intermediate users who have also shared the post, and
are scaled according to estimated influence.

\section{Conclusion}
We have outlined the structure of an information visualization
platform for the exploration of communication networks on the Twitter
microblogging platform. We motivate our design decisions in terms of
traditional benchmarks for effective interface design as well as in terms of
insights gleaned from close collaboration with journalism scholars actively
engaged in research on the role of social media and the public sphere. The
product of this collaboration is a visual analytics tool that allows
non-technical users to explore the results of large-scale computational and
statistical analyses in an intuitive and informative fashion. 

As increasing amounts of data on online discourse and deliberation are
collected, we see a rising demand for technologies that make these data
intelligible to the broader research community.  We envision that systems which
leverage in equal measure the strengths of the computational and social
sciences will act as one of the primary drivers of research innovation for
years to come. Work continues on the Truthy system, and in the near future, we
will extend the visualizations with temporal and geographic information.  The
visualizations described in this paper are all currently available on the
Truthy website, and we encourage readers to explore them.

{\small \paragraph{Acknowledgements.} We are grateful to Emily Metzgar and Hans Ibold for their invaluable insights and support. This work is supported in part by NSF REU award CCF-1101743.} 

\bibliography{visualizations}{}
\bibliographystyle{plain}
\end{document}